\documentstyle[12pt]{article}
\begin{document}
\title{Exact solutions of effective mass Schr\"{o}dinger equations}
\author{B. Roy\thanks{E-mail : barnana@isical.ac.in} ~ and ~
P. Roy\thanks{E-mail : pinaki@isical.ac.in}\\ Physics \& Applied
Mathematics Unit\\ Indian Statistical Institute \\ Calcutta  700035\\ India}
\maketitle

\begin{abstract}
We outline a general method of obtaining exact solutions of Schr\"{o}dinger equations
with a position dependent effective mass. Exact solutions of several potentials
including the shape invariant potentials have also been obtained. 


\end{abstract}

{\bf 1.} Apart from being an interesting topic itself Schr\"{o}dinger equations
with a position dependent effective mass have found 
wide applications in the study of electronic properties of semiconductors \cite{bastard},
quantum dots \cite{sera}, quantum liquids \cite{arias} etc. Although exact solutions of
are difficult to obtain, for certain potentials the effective mass Schr\"{o}dinger
equation can be solved \cite{dekar,dutra}. Recently supersymmetric techniques
have also been applied to obtain a few exactly solvable 
potentials \cite{plas,milan}. In this article we shall use a simple method ( which
depends on the equivalence of second order differential operators \cite{kamran})
to obtain exact solutions of Schr\"{o}dinger equation with a position dependent
mass. In particular we shall obtain exact solutions of a number of hitherto
unknown potentials. To demonstrate the simplicity of the method we shall also 
rederive the solutions of some of the potentials studied earlier \cite{plas}.

{\bf 2.} When the mass depends on the position the kinetic energy can be defined in
several ways. Here we shall be following L\'{e}vy-Leblond \cite{levy} and in this
case the Schr\"{o}dinger equation is given by
\begin{equation}
-\frac{d}{dx}(\frac{1}{2m(x)}\frac{d\psi(x)}{dx}) + V(x)\psi(x) = E\psi(x) \label{sch1}
\end{equation}   
The wave function $\psi(x)$ should be continuous at the mass discontinuity and
the derivative of the wave function should satisfy the following condition:
\begin{equation}
\frac{1}{m(x)}\frac{d\psi(x)}{dx}|_- = \frac{1}{m(x)}\frac{d\psi(x)}{dx}|_+
\end{equation}
We now perform the transformation 
\begin{equation}
\psi(x) = [2m(x)]^{\frac{1}{4}}\phi(x) \label{t1}
\end{equation}
on equation (\ref{sch1}) and obtain
\begin{equation}
-\frac{1}{2m(x)}\phi''(x) + \frac{1}{4}(\frac{m'(x)}{m^2(x)})\phi'(x) + [\frac{4m''(x)-7{m'}^2(x)}{32m^3(x)}] \phi(x) + V(x)\phi(x) = E\phi(x) \label{sch2}
\end{equation}
where the prime indicates differenciation with respect to $x$. Next we make a
change of the independent variable defined by
\begin{equation}
{\bar x} = \int^x \sqrt{2m(y)} dy \label{t2}
\end{equation}
Using (\ref{t2}) in equation (\ref{sch2}) we get
\begin{equation}
-\frac{d^2\phi(\bar x)}{d{\bar x}^2} + \Omega(\bar x)\phi(\bar x) = E\phi(\bar x) \label{sch3}
\end{equation}
where for the sake of convenience we have used $\phi(x)|_{x=\bar x} = \phi(\bar x)$
and $\Omega(\bar x)$ is defined by
\begin{equation}
\Omega(\bar x) = V(x=\bar x) + [\frac{4m''(x)-7{m'}^2(x)}{32m^3(x)}]_{x=\bar x} = V(x=\bar x)+V_1(x=\bar x)\label{omega}
\end{equation} 
It is important to note that the change of variable in (\ref{t2}) may not always
be invertible or at least not easily invertible. But this does not really pose
a problem as far as solvability of (\ref{sch1}) is concerned. This is because 
$\bar x$ as a function of $x$ is explicitly known from (\ref{t2}) and if we choose $V(x)$ such that
\begin{equation}
V(x) = V_2(\bar x) - V_1(\bar x)
\end{equation}
where $V_2(\bar x)$ is a solvable potential then the spectrum of (\ref{sch3})
will be known and this in turn will give us the spectrum of (\ref{sch1}). The
corresponding wave functions can be obtained using (\ref{t1}). In the next
section we illustrate the method with a few examples.

{\bf 3.} In order to deal with specific potentials it is necessary to prescribe the
mass function $m(x)$. First we consider the choice used in \cite{plas}:
\begin{equation}
m(x) = (\frac{\alpha+x^2}{1+x^2})^2~~~,~~~m(0) = \alpha^2~~~,~~~m(\pm\infty) = 1 \label{m1}
\end{equation} 
Then form (\ref{t2}) we get
\begin{equation}
\bar x = \sqrt 2[x+(\alpha-1)tan^{-1}x]~~,~~-\infty<\bar x<\infty \label{barx}
\end{equation}
and using (\ref{barx}) in (\ref{omega}) we find
\begin{equation}
V_1(x) = \frac{(\alpha-1)}{2(\alpha+x^2)^4}[-3x^4+(2\alpha-4)x^2+\alpha]
\end{equation}
As we mentioned in the last section it is now necessary to choose $V_2(\bar x)$
to be a solvable potential. Let us first consider the harmonic oscillator:
\begin{equation}
V_2(\bar x) = \frac{1}{4}{\bar x}^2 \label{v21}
\end{equation}
so that
\begin{equation}
E_n = (n+\frac{1}{2}) \label{ener1}
\end{equation} 
Then the original potential $V(x)$ is given by
\begin{equation}
V(x) = \frac{1}{2}[x+(\alpha-1)tan^{-1}x]^2 + \frac{(\alpha-1)}{2(\alpha+x^2)^4}[3x^4+(4-2\alpha)x^2-\alpha] \label{p1}
\end{equation}
The potential (\ref{p1}) has harmonic oscillator spectrum given by (\ref{ener1}) and using (\ref{t1}) and (\ref{t2})
the wave functions are found to be
\begin{equation}
\psi_n(x)=N_n\sqrt{\frac{\alpha+x^2}{1+x^2}}exp(-{\bar x}^2/2)H_n(2^{-\frac{1}{4}}\bar x)
\end{equation}
where $\bar x$ is given by (\ref{barx}). We note that the potential (\ref{p1})
is a shape invariant potential \cite{plas}. Next we choose $V_2(\bar x)$ as the Morse potential
\begin{equation}
V_2(\bar x) = {\lambda}^2[1-exp(-x)]^2~~,~~E_n = [2\lambda(n+\frac{1}{2})-(n+\frac{1}{2})^2]~~,~~0\leq n \leq[\lambda-1] \label{morse}
\end{equation}
Then for $V(x)$ we get the following potential which has the same spectrum (\ref{morse})
as the Morse potential:
\begin{equation}
V(x) = {\lambda^2}\left\{1-exp[-(x+(\alpha-1)tan^{-1}x)]\right\}^2 + \frac{(\alpha-1)}{2(\alpha+x^2)^4}[3x^4+(4-2\alpha)x^2-\alpha] \label{p2}
\end{equation}

As another example let us consider the soliton potential
\begin{equation}
V_2(\bar x) = -\lambda(\lambda+1)sech^2\bar x \label{soliton}
\end{equation}
with energy 
\begin{equation}
E_n = -(\lambda-n)^2~~,~~n<\lambda \label{e3}
\end{equation}
The initial potential is then given by
\begin{equation}
V(x) = -\lambda(\lambda+1)sech^2[{\sqrt 2}(x+(\alpha-1)tan^{-1}x)] + \frac{(\alpha-1)}{2(\alpha+x^2)^4}[3x^4+(4-2\alpha)x^2-\alpha] \label{p3}
\end{equation}
with energy given by (\ref{e3}). 

Finally we consider an important class of potential,namely,quasi exactly solvable
potential \cite{turbiner} for which a part of the spectrum can be determined
analytically. A typical representative potential of this class is given by \cite{turbiner}
\begin{equation}
V(\bar x) = {\bar x}^6 - (8j+3){\bar x}^2~~,~~j=0,\frac{1}{2},1,... \label{qes}
\end{equation}
and for (\ref{qes}) $(2j+1)$ levels can be determined exactly. For example for
$j=\frac{1}{2}$ two levels $E_{\mp}$ can be found: 
\begin{equation}
E_{\mp} = \mp 2\sqrt 2 \label{qes1}
\end{equation}
The quasi exactly solvable potential $V(x)$ corresponding to (\ref{sch1}) is given by
\begin{equation}
V(x) = [x+(\alpha-1)tan^{-1}x]^6 - 7[x+(\alpha-1)tan^{-1}x]^2 + \frac{(\alpha-1)}{2(\alpha+x^2)^4}[3x^4+(4-2\alpha)x^2-\alpha] \label{p4}
\end{equation}
and its eigenvalues are given by (\ref{qes1}). We note that wave functions
corresponding to the potentials (\ref{p2}),(\ref{p3}) and (\ref{p4}) can be
obtained using (\ref{t1}) from those of (\ref{morse}),(\ref{soliton}) and 
(\ref{qes}) respectively. It is thus clear that for every potential for which
spectral properties of the constant mass Schr\"{o}dinger equation are known
we can obtain a corresponding potential for the effective mass Schr\"{o}dinger equation with identical spectral properties.

{\bf 4.} The problem of isospectrality of Hamiltonians (when the mass is constant)
has been thoroughly investigated \cite{khare}. When the mass depends on the
space coordinate has also been investigated using supersymmetry \cite{milan}.
Here we shall demonstrate that two Hamiltonians can indeed be isospectral even
when both the mass and the potential are different. To show this we choose a mass
of the form
\begin{equation}
m(x) = (\frac{\alpha+x^2}{1+x^2})^4 \label{m2}
\end{equation}
Then from (\ref{t1}) we get
\begin{equation}
\bar{\bar x} = 2\sqrt 2[2x+\frac{(\alpha-1)^2x}{(1+x^2)}+(\alpha-1)(\alpha-3)tan^{-1}x]~~,-\infty<\bar{\bar x}<\infty
\end{equation}
while (\ref{omega}) gives us
\begin{equation}
V_1(x) = \frac{(\alpha-1)(1+x^2)^2}{(\alpha+x^2)^6}[-3x^4+(5\alpha-7)x^2+\alpha] 
\end{equation} 
Now for $V_2(\bar{\bar x})$ we take
\begin{equation}
V_2(\bar{\bar x}) = \frac{{\bar{\bar x}}^2}{4} \label{v23}
\end{equation}
so that its spectrum is given by (\ref{ener1}).
In this case $V(x)$ is given by
\begin{equation}
V(x) = \frac{1}{8}[2x+\frac{(\alpha-1)^2x}{(1+x^2)}+(\alpha-1)(\alpha-3)tan^{-1}x]^2 + \frac{(\alpha-1)(1+x^2)^2}{(\alpha+x^2)^6}[3x^4+(7-5\alpha)x^2-\alpha] \label{p5}
\end{equation}
Since (\ref{v23}) is a harmonic oscillator potential it follows that the effective
mass Schr\"{o}dinger equations for the potentials (\ref{p1}) and (\ref{p5}) with
masses given by (\ref{m1}) and (\ref{m2}) respectively are exactly isospectral.

{\bf 5.} In this article we have described a general method of obtaining exact solutions
of effective mass Schr\"{o}dinger equation. In particular the method has been
applied to obtain solutions of several potentials which have so far been unknown.
It has also been shown that Schr\"{o}dinger equations with
different effective masses as well as different potentials can be isospectral.

\end{document}